\documentstyle[aps,preprint,amssymb]{revtex}

\begin{document}
\draft

\title{Jamming creep of a frictional interface}

\author{L. Bureau, T. Baumberger, C. Caroli}

\address{Groupe de Physique des Solides, Universit\'es Paris 6 et 
7, UMR CNRS 7588, 2 place Jussieu, 75251 Paris, Cedex 05, France}

\date{\today}
\maketitle

\abstract{
We measure the displacement response of a frictional multicontact 
interface between identical polymer glasses
to a biased shear force oscillation. We evidence the existence, for 
maximum forces close below the nominal static threshold, of a 
jamming creep regime governed by an ageing-rejuvenation competition 
acting within 
the micrometer-sized contacting asperities. Quantitative analysis of the 
creep curves suggests that another such mechanism might be at work 
within the nanometer-thick adhesive junctions.
}


Solid friction between two macroscopic solids is commonly 
characterized in terms of {\it (i)} a static force threshold below 
which no relative displacement is supposed to take place, {\it (ii)} 
a dynamic friction coefficient, measured in stationary motion. 
However, recent experiments performed on
multicontact 
interfaces (MCI), {\it i.e.} interfaces between two solids with
rough surfaces, pressed together under a load $N$, have revealed that 
\cite{Elast}:\\
{\it (i)} for shear forces $F$ such that $F/N\ll\mu_s$, where $\mu_s$ is the 
static threshold, the pinned interface responds elastically, {\it via} 
the reversible
deformation of the contacting asperities,\\
{\it (ii)} for $F\lesssim \mu_s N$, creeplike irreversible sliding 
is observed.

Clearly, the study of the latter regime of incipient sliding should give 
access to precise information about the underlying pinning/depinning 
dynamics.

It is now well established \cite{PRB} that, for a MCI, the variations 
of the friction force $F=\sigma_{s}\Sigma_{r}$ (with $\Sigma_{r}$ the 
real area of contact, and $\sigma_{s}$ the interfacial shear stress) 
are governed, at low velocities, by the competition between two 
effects: {\it (i)} an age strengthening effect resulting from the 
logarithmic creep growth,
under the high (geometry-enhanced) normal load, of the sparse contacts between 
load bearing asperities. When motion
starts, contacts get gradually destroyed, after a lifetime or age 
$\Phi$, and replaced by fresh ones. So, while the interface sits 
still, it ages (strengthens), when it slides, it rejuvenates (weakens). 
Full refreshment occurs, on average, after sliding a micrometric memory 
length $D_0$. {\it (ii)} Weakening when 
sliding is counteracted by the velocity-strengthening 
interface rheology: $\sigma_{s}(\dot{x})=\sigma_{s0}\left[1+\alpha 
\ln(\dot{x}/V_{0})\right]$, with $\dot{x}$ the instantaneous sliding 
speed. It results from thermally activated premature depinning events 
within the 
nanometer-thick adhesive junctions between contacting asperities. 

Both effects yield logarithmic variations of $F$. One thus expects 
creep to exhibit a strong, exponential, sensitivity to 
forces close to the nominal static threshold.

Now, the experiments reported in \cite{Elast} were performed 
under static loading through a spring of finite stiffness. As such, they 
did not provide a force control fine enough to study incipient creep 
accurately. So, we have chosen, in this work, to 
probe it {\it via} the response to a biased {\it a.c.} shear force. With a 
bias $F_{dc}\ll \mu_{s}N$, and an amplitude such that the maximum 
force $F_{max}$ lies in the tangential creep range, the interface 
should experience, during each oscillation period, an alternation of two regimes:
{\it (i)} for $F$ close to $F_{max}$, a sliding phase during which 
rejuvenation is at work, yielding an age variation 
$\Delta \Phi_{slide}$, and {\it (ii)} as $F$ decreases, the slip 
velocity decreases quasi-exponentially, and the slider enters, for the rest of 
the period, a quasi static phase where age grows 
linearly with time by an amount $\Delta 
\Phi_{stat}$.
Such a competition between rejuvenation and ageing is akin to that 
invoked to model soft glassy rheology (SGR) \cite{SGR}. 

If, say, $\Delta \Phi_{slide}+\Delta \Phi_{stat}<0$, the interface 
will weaken, leading to a larger slip during the next period, etc. 
One thus expects the dynamics to bifurcate between self-accelerated 
unlimited slip and self-decelerated, jamming \cite{Jam}, creep. The experiments 
reported below fully confirm this qualitative scenario. Moreover,
quantitative analysis of the data, based upon the Rice-Ruina (RR) model 
\cite{RR}, allows us to show 
that the rejuvenation-ageing process cannot be fully ascribed to 
variations of $\Sigma_{r}$, leading us to conclude that the interfacial 
rheology is most likely, itself, of the SGR type.

    {\it Experiments.---} The experimental setup has been 
    fully described in \cite{RSI}; it is sketched in the inset of 
    Fig.\ref{fig:1}a. 
    Two poly(methyl methacrylate) (PMMA)
    samples, with lapped surfaces of roughness $R_{a}=1\, \mu$m, are glued on a 
    slider and a track and form the multicontact interface. The 
    slider, of nominal area $4$ cm$^{2}$, 
    rests on the track, inclined at $\theta=20^{\circ}$ from the
    horizontal. The tangential ($F_{dc}$) to normal ($N$) load 
    ratio $\gamma_{dc}=\tan \theta=0.36$ is well 
    below the static threshold $\mu_{s}\approx 0.6$ (see \cite{PRB}) 
    and no sliding 
    occurs. Imposing a harmonic motion to the track then results in 
    an inertial shear loading of the slider, of amplitude 
    $F_{ac}=\gamma_{ac}N$, with $\gamma_{ac}\leq 0.5$. The 
    frequency $f=80$ Hz is chosen well below the natural frequency of 
    the slider-interface system, $f_{0}=800$~Hz, so that the inertia 
    associated to its relative motion in the track frame can be 
    neglected.
    We measure the displacement $X$ of the center of mass of the 
    slider by means of a 
    capacitive displacement gauge, with a noise amplitude 1 nm over 
    its whole 0--500 Hz bandwidth.
    In order to prepare the system in as reproducible as possible an 
    initial state, the slider is placed on the track and a large 
    $\gamma_{ac}$ is then imposed, in order to make it slide a few micrometers 
    in the 
    direction of $F_{dc}$. The harmonic force is then suddenly 
    stopped, which results in an 
    elastic recoil of the contacting asperities \cite{CaNo}. 
    This method reduces the relative dispersion on 
    interfacial 
    stiffness  values (see \cite{RSI}) to only $10\%$. This value agrees with 
    the expected statistical dispersion due to the finite number of 
    load bearing contacts, which we can estimate to be of order 50.
    A time $t_{wait}$ is then waited before reswitching the 
    harmonic shear loading, either as a linear ramp of 
    amplitude, 
    until gross 
    sliding occurs, or as a step with rising time $\leq 
    0.1$~s. 
    
    {\it Results.---}The displacement response $X(t)$ of the slider 
    to a ramp 
    $\gamma_{ac}(t)$ is illustrated on Fig.\ref{fig:1}a. Also shown is the 
    average displacement $X_{dc}$ measured by filtering $X(t)$ 
    through a low-pass filter of cutoff frequency 8 Hz.
    In region (I), the slider oscillates about a constant average 
    position, no irreversible slip occurs: the MCI responds elastically.
    Region (III) 
    corresponds to accelerated sliding. In region (II) between these two regimes,
    the average displacement 
    $X_{dc}$, {\it i.e.} the slipped distance, increases continuously. 
    We use this ramp test to define a threshold 
    $\gamma_{s}=(F_{dc}+F_{ac})/N$ 
    such that the average sliding 
    velocity $dX_{dc}/dt=100\, \mu$m.s$^{-1}$. 
    We thus obtain, for 
    $t_{wait}=600$~s and $\dot{\gamma}_{ac}=0.1$~s$^{-1}$, 
    $\gamma_{s}=0.59\pm 0.03$. The scattering, of order 10\%, 
    is consistent with that of the stiffness.
    
     One can see, on Fig.\ref{fig:1}a, that the intermediate creep regime (II) 
     corresponds to a narrow range of 
     $\gamma_{max}=\gamma_{dc}+\gamma_{ac}$. Slow 
     creep is studied by choosing a value $\gamma_{max}$ in this 
     range, setting, at $t=t_{wait}$, the amplitude stepwise to 
     $\gamma_{ac}$, and recording $X_{dc}(t)$.
    
   The set of creep curves displayed on Fig.\ref{fig:1}b all correspond to 
   $t_{wait}=300$ s, $\gamma_{max}=0.54$. The large dispersion 
   between $X_{dc}$ for various runs must therefore result from the 
   statistical dispersion of the MCI initial state. After slipping by 
   a finite amount, of order 10--100 nm, over the rising time (0.1 s) of $\gamma_{ac}$, the 
   slider performs a slowly self decelerating creep. After, typically, 
   $10^{4}$ s, the 
   slip velocity has decreased to non-measurable values, indicating 
   a saturating, jamming dynamics. We attribute the large 
   dispersion of the creep curves to the expected above-mentioned 
   exponential sensitivity of the dynamics to 
   $\gamma_{s}-\gamma_{max}$. A direct confirmation of this is 
   obtained from the experiment presented on Fig.\ref{fig:1}c: it shows that a 
   3\% step of $\gamma_{max}$ turns quasi-jamming into accelerated 
   sliding.
   
   These ideas can be checked in a more quantitative way as follows: 
   as long as the age of the MCI has not been appreciably modified by 
   the creeping dynamics itself, we expect the characteristic time 
    for creep, $t_{c}$, to be that for thermally activated depinning of a 
   typical nm$^{3}$ pinned unit within the adhesive layer: 
   $\ln(t_{c})\sim Cst + \Delta E/kT$, where 
   $\Delta E$ is the energy barrier to be jumped by an element under 
   reduced load $\gamma_{max}$. Close to the depinning threshold 
   \cite{Bo} $\Delta E/kT\sim (\gamma_{s}-\gamma_{max})/A$. The RR rate 
   parameter $A=\alpha \sigma_{s0}/\bar{p}$, with $\bar{p}$ the 
   average pressure on the microcontacts, has been measured, for 
   PMMA/PMMA, to be 0.013 \cite{PREvibro}. Hence, we expect the slipped 
   distance in run (i) to scale as $X_{dc}^{i}(t)=\tilde{X}(t/C_{i})$, 
   where $\ln(C_{i})\sim (\gamma^{i}_{s}-\gamma_{max})/A$. Fig.\ref{fig:2} 
   shows the 
   set of creep curves resulting from such a scaling. Indeed, the 
   collapse onto a master curve is very good in the short time range. 
   Moreover, we find the maximum spread of the scaling factors $\max 
   \vert \ln(C_{i}/C_{j}) \vert \approx 4.4$, in excellent agreement with $\max 
   \vert \Delta (\gamma^{i}_{s}-\gamma_{max}) \vert /A \approx 4.6$.
   
   {\it Discussion.---} We now analyze our results within the framework 
   of the Rice-Ruina phenomenology \cite{RR}, which has proved to account very 
   well for the stick-slip frictional dynamics of MCIs \cite{PRB}.
   It models the above described rejuvenation-ageing process and the 
   velocity strengthening interface rheology as follows. 
   
   The friction 
   coefficient reads:
   \begin{equation}
       \mu = F/N=
       \mu_{0}+A\ln\left(\frac{\dot{x}}{V_{0}}\right)
       +B\ln\left(\frac{\Phi V_{0}}{D_{0}}\right)
   \label{eq:RR1}
   \end{equation}
   $\Phi$ is the interface age, 
   $\mu_{0}$ 
   the friction coefficient at reference velocity $V_{0}$. The  
   instantaneous interfacial sliding velocity $\dot{x}$ is related 
   to the center of mass position by $\dot{x}=d(X-F/\kappa)/dt$, 
   with $\kappa$ the interfacial elastic stiffness \cite{PREvibro}.
   
   The  age $\Phi$ evolves according to:
   \begin{equation}
       \label{eq:RR2}
       \dot{\Phi}=1-\frac{\dot{x}\Phi}{D_{0}}
   \end{equation}
   On the r.h.s. of Eq.(\ref{eq:RR2}), the first and second 
   terms correspond, respectively, to time ageing and slip rejuvenation.
   
   We have performed numerical integrations of this set of 
   differential equations to calculate the slipped distance 
   $X_{dc}(t)$ with, in Eq.(\ref{eq:RR1}), 
   $F/N=\gamma_{dc}+\gamma_{ac}\cos(\omega t)$, and initial conditions
   for slip and age 
   $X_{dc}(0)=0$, $\Phi(0)=t_{wait}=300$~s.
   We have used for the 
   memory length $D_{0}$ and the dimensionless parameters $A$ and $B$ 
   the values $D_{0}=0.42\, \mu$m, $A=0.013$, $B=0.026$, obtained from
   previous measurements on PMMA \cite{PREvibro}. We choose 
   $V_{0}=1\, \mu$m.s$^{-1}$. Due to the exponential amplification by 
   the creep dynamics of the small variations of the absolute 
   friction level between various runs, $\mu_{0}$ must be left free. 
   This unique fitting parameter is tuned so as to adjust the 
   calculated and measured values of $X_{dc}$ at the end of the run.
   A typical example of such fits is
   shown on Fig.\ref{fig:3}. It yields  
   $\mu_{0}=0.42865$\cite{footnote}, fully compatible with previous data \cite{PRB}.
   
   These fits appear to be very good  at 
   long times $t\gtrsim 1000$~s, 
   that is in the quasi-jammed regime where static ageing becomes 
   dominant. The corresponding asymptotic dynamics can be analyzed directly. From 
   Eq.(\ref{eq:RR1}):
   \begin{equation}
       \frac{\dot{x}}{V_{0}}=\left(\frac{D_{0}}{V_{0}\Phi}\right)^{\beta}
       \exp\left[{\frac{\gamma_{max}-\mu_{0}}{A}}\right]
       \exp\left[{\frac{\gamma_{ac}(\cos(
       \omega t)-1)}{A}}\right]
   \label{eq:xpoint}
   \end{equation}
   with $\beta=B/A$.
   
   Let us consider the oscillation period centered at $t_{1}$, with 
   $\gamma(t_{1})=\gamma_{max}$. The velocity $\dot{x}$ only takes on significant 
   values for $\gamma \approx \gamma_{max}$, {\it i.e.} for $\vert 
   \tau \vert=\vert t-t_{1} \vert \ll T$. So, in Eq.(\ref{eq:xpoint}), 
    $\cos(\omega 
   t)-1\approx -\omega^{2}\tau^{2}/2$, and:
   \begin{equation}
       \frac{\dot{x}}{V_{0}}\approx 
       \left(\frac{\Phi_{c}}{\Phi}\right)^{\beta}\exp\left[
       -\frac{\tau^{2}}{\tau_{c}^{2}}\right]
   \label{eq:xpointapprox}
   \end{equation}
   with the constant
   $\Phi_{c}^{\beta}=(D_{0}/V_{0})^{\beta}\exp\left[(\gamma_{max}-\mu_{0})/A\right]$ and 
   $\omega \tau_{c}=(2A/\gamma_{ac})^{1/2}$.
   
   To lowest order, the slip-induced $\Phi$ variation can be 
   neglected, and Eq.(\ref{eq:RR2}) yields:
   \begin{equation}
       \Phi\approx t+\Theta
   \label{eq:phistat}
   \end{equation}
   where the integration constant $\Theta$ is the age at some 
   ``initial'' time within the quasi-jammed regime. 
   
   From Eqs.(\ref{eq:xpointapprox}) and (\ref{eq:phistat}), the 
   increment of slip over this period:
   \begin{equation}
       \Delta x\approx \int_{-\infty}^{+\infty}
       \frac{\Phi_{c}^{\beta}V_{0}}{(t_{1}+\tau+\Theta)^{\beta}}\exp\left[
       -\frac{\tau^{2}}{\tau_{c}^{2}}\right]d\tau 
   \end{equation}
   where the integration can be extended to infinity due to the
   gaussian decay of $\dot{x}$. Then, for $t_{1}\gg \Theta$:
   \begin{equation}
       \Delta x
       \approx 
       \frac{V_{0}\Phi_{c}^{\beta}\sqrt{\pi}\tau_{c}}{t_{1}^{\beta}}+O\left(
       \frac{1}{t_{1}^{\beta +1}}\right)
   \label{eq:Deltax}
   \end{equation}
   From this, the sliding velocity $\dot{X}_{dc}$, coarse-grained 
   over the period $T$:
   \begin{equation}
       \dot{X}_{dc}(t)\approx \frac{\Delta x}{T}\approx 
       V_{0}\sqrt{\frac{A}{2\pi\gamma_{ac}}}
       \frac{\Phi_{c}^{\beta}}{t^{\beta}}
   \label{eq:Vmoy}
   \end{equation}
   and we obtain for the slipped distance:
   \begin{equation}
       X_{dc}(t)\approx Cst-\frac{V_{0}}{\beta -1}
       \sqrt{\frac{A}{2\pi\gamma_{ac}}}
       \frac{\Phi_{c}^{\beta}}{t^{\beta -1}}
   \label{eq:slip}
   \end{equation}
   Since, for our system, $\beta=2$, we thus expect the creeped distance to 
   approach its saturation level 
   as $1/t$. 
   
   Our experimental data are seen on the insert of Fig.\ref{fig:3} to 
   fit the jamming asymptotics predicted by the RR model with 
   excellent accuracy.
   
   However, for all the experimental runs, we find (see 
   Fig.\ref{fig:3}) that, although the overall shape of $X_{dc}(t)$ is 
   reasonably well described by the RR fits, the agreement is clearly 
   not quantitative at short times ($t\lesssim 1000$~s). In this time 
   bracket, the RR model systematically underestimates $X_{dc}(t)$, hence the 
   rejuvenation efficiency of slip. In particular, it by no 
   means accounts for the fast increase of $X_{dc}$, which amounts 
   typically to $\sim 10$\% of the total slip, occurring over 
   the stepping time.
   
   This strongly hints at the fact that the RR model, while it 
   very well describes established sliding, misses some important 
   feature of incipient sliding. We suspect that this missing feature 
   might be slip-induced rejuvenation, {\it i.e.} dynamical weakening 
   of $\sigma_{s}$, within the nanometer-thick adhesive junctions 
   themselves. Indeed, these certainly have an amorphous solid 
   structure when pinned, and they flow beyond a stress threshold. As 
   such, they can reasonably be expected to behave as soft glassy 
   materials, whose rheology is now interpreted \cite{SGR} in terms of structural 
   ageing/rejuvenation competition. If this turns out to be the case, 
   two such mechanisms would be at work in the MCI solid friction, on 
   the two scales of, respectively, the micrometric asperities, and 
   the nanometric pinning units.
   
   This issue is, in particular, of primary relevance to the 
   modelling of the dynamics of interfacial shear fracture 
   \cite{RiCo,Frac}.
   We plan to investigate it by studying the frictional 
   dynamics of rough PMMA sliding over smooth hard glass. Since, with 
   such a system, the microcontact population is unaffected by motion, 
   rejuvenation, if observed, will have to originate from the adhesive 
   junctions.
   


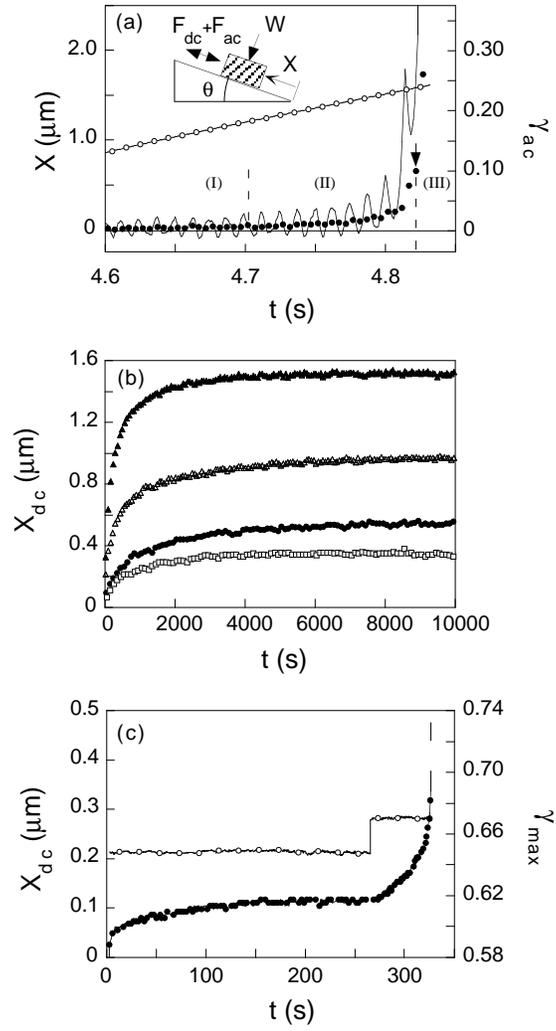
    \begin{figure}
	\caption{(a) Instantaneous (line) and averaged ($\bullet$) 
	displacement response of the slider to a biased oscillating shear 
	force of ramped reduced amplitude $\gamma_{ac}(t)$ ($\circ$). 
	The bias $\gamma_{dc}=0.36$.
	The arrow indicates the point at which the averaged velocity reaches
	$100\, 
	\mu$m.s$^{-1}$. Inset: experimental setup. (b) Recordings of the 
	average creep displacement $X_{dc}$ for four runs performed under identical nominal 
	conditions: $\gamma_{dc}=0.36$, $\gamma_{ac}=0.18$, 
	$t_{wait}=300$~s. The wide scattering of the curves results from 
	dynamical amplification of the statistical dispersion of the 
	interfacial strength.
	(c) Transition from jamming to unbounded slip ($\bullet$) triggered 
	by a 3\% jump of $\gamma_{max}=\gamma_{dc}+\gamma_{ac}$ ($\circ$).}
	\label{fig:1}
    \end{figure}
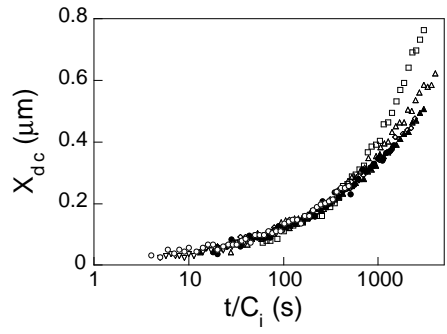
    \begin{figure}
        \caption{Scaled plot of 7 creep curves (same 
        nominal conditions as for Fig.1b). The reference run ($C=1$) 
        corresponds to ($\circ$) symbols.}
        \label{fig:2}
    \end{figure}
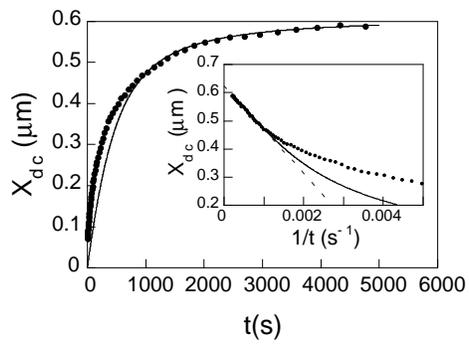
    \begin{figure}
        \caption{An experimental creep curve ($\bullet$) and 
        its fit according to RR model (line). With $\gamma_{dc}=0.36$, 
        $\gamma_{ac}=0.18$, 
	$t_{wait}=300$~s. Inset: the same plotted {\it 
        versus} $1/t$, the dashed line indicates the quasi-jamming asymptotics 
       (see text).}
       \label{fig:3}
   \end{figure}
   

\begin{references}
\bibitem{footnote} The five decimal places are necessary to adjust 
	the final $X_{dc}$ to 1\%.
\end{references}
\end{document}